\documentclass[12pt]{iopart}
 \expandafter\let\csname equation*\endcsname\relax
  \expandafter\let\csname endequation*\endcsname\relax
\usepackage{amsmath}
\usepackage{amsfonts}
\usepackage{mathrsfs}
\usepackage{graphicx}
\usepackage[colorlinks={true}]{hyperref}
\hypersetup{citecolor={blue}, filecolor={blue}, linkcolor={blue}, urlcolor={blue}}
\usepackage{color}
\usepackage{txfonts}

\begin{document}
\title{Achieving robust and high-fidelity  quantum control via spectral phase optimization}
\author{Yu Guo$^{1,2}$, Daoyi Dong$^{1}$, and Chuan-Cun Shu$^{1}$}
\address{$^{1}$School of Engineering and Information Technology, University of New South Wales, Canberra, Australian Capital Territory 2600, Australia}
\address{$^{2}$School of Physics and Electronic Science, Changsha University of Science and Technology, Changsha  410114, China}
\ead{c.shu@unsw.edu.au; chuancunshu@gmail.com}

\begin{abstract}
Achieving high-fidelity control of quantum systems is of fundamental importance in physics, chemistry and quantum information sciences.
However, the successful implementation of a high-fidelity quantum control scheme also requires robustness against control field fluctuations. Here, we demonstrate a robust optimization method for control of quantum systems by optimizing
the spectral phase of an \emph{ultrafast} laser pulse, which is  accomplished in the framework of frequency domain quantum optimal control theory.
By incorporating  a filtering function of frequency into the optimization algorithm, our numerical simulations in an abstract two-level quantum system as well as  in a three-level atomic rubidium show that  the optimization procedure  can be enforced  to search optimal solutions  while achieving remarkable robustness against the control field fluctuations, providing an efficient approach to  optimize the spectral phase of the ultrafast  laser pulse to achieve a desired final quantum state of the system.

\end{abstract}

 \section{Introduction}
In spite of rapid progress in quantum technology \cite{nature:541:9},  it remains a challenge to achieve high-fidelity quantum control while tolerating uncertainty and quantum decoherence effects \cite{nature:540:44,ejpd}. In regard to the latter, the advanced ultrafast (pico-, femto- and attosecond) laser technique provides an alternative approach to control quantum dynamical processes on extremely short time scales before quantum decoherence effect plays roles, leading to a variety of applications from science to industry \cite{arpc:60:277,nc:7:11200,nc:8:238,jpcl:8:1,jpcl:8:2229}.  Mathematical description of an ultrafast laser pulse can be written as
 \begin{eqnarray}
\mathcal{E}(t)=\mathrm{Re}\Bigg[\int_{0}^{\infty}\boldsymbol{E}(\omega)\exp(-i\omega t)d\omega\Bigg], \label{phase-only}
\end{eqnarray}
in terms of a complex function $\boldsymbol{E}(\omega)$ in the frequency domain, where $\boldsymbol{E}(\omega)=\mathcal{A}(\omega)\exp(i\phi(\omega))$ is a product of
a spectral amplitude $\mathcal{A}(\omega)$ and a spectral  phase
$\phi(\omega)$. To achieve high-fidelity of quantum control with optimal ultrafast laser pulses, quantum optimal control experiment (QOCE) relying on the computational intelligence of an evolutionary algorithm is often employed to find the ``best" combination of $\mathcal{A}(\omega)$ and $\phi(\omega)$ in the frequency domain \cite{science:282:919,sience:299:536,pccp:9:2470,prl:112:143001}, whereas many quantum optimal control theory (QOCT) methods used for simulations are accomplished in the time domain by directly shaping the temporal field $\mathcal{E}(t)$ \cite{ejpd,jpb:4:R175,heshel:njp}. It is clear that there is a big mismatch between theory and experiment.
Recently, a general frequency domain quantum optimal control theory (FDQOCT) was established \cite{pra:93:033417,pra:inpress}, which can be utilized in a monotonic convergence fashion to  optimize the spectral field $\boldsymbol{E}(\omega)$ while incorporating  multiple internal limitations as well as external constraints into the optimization algorithm.
 Provided that the involved limitations and constraints are sufferable, the broad successes of QOCT and QOCE applications have shown that there usually exist
many solutions capable of achieving the same
observable value \cite{jcp:134:194106,jcp:137:134113,pra:91:043401,pra:95:063418}. A question of particular interest  is how  to search optimal robust solutions against various uncertainties. Many efforts have been put forth to address this question even for the simplest quantum system of two states \cite{A1,A2,A3,A4,Automatica,np:10:825,pra:89:023402,prl:111:050404,prl:100:103004,prl:106:233001,pra:93:043419,pra:95:062325,scirep}. Optimization approaches usually are accomplished  in the time domain  by solving a robust optimization problem \cite{pra:49:2241,mrm:36:401,prl:88:052326,arXiv:1704.07653v1,pra:96:022309}, which  have intuitive applications  with state-of-the-art technologies for microsecond radio-frequency pulses.
\\ \indent
In this paper, we present a theoretical investigation  in the framework of FDQOCT for optimal robust control of quantum systems  without taking the uncertainty into account. To highlight quantum coherence effects while reducing  the ``search space"  \cite{jpcl:6:4032,jcp:144:141102,jpcl:6:824,prl:86:47,jpcl:3:2458,jcp:136:044303}, the FDQOCT is specifically employed to shape the spectral phase $\phi(\omega)$ of an \emph{ultrafast} nonadiabatic  pulse by keeping the spectral amplitude $\mathcal{A}(\omega)$ unchanged, which leads  to the spectral-phase-only optimization (SPOO) algorithm. To illustrate this method without complexities from the dimension
of systems, we first employ the SPOO algorithm to obtain a complete population inversion between two levels, for which previous theoretical and experimental studies  have demonstrated  that a quadratic spectral phase function of $\phi(\omega)=\beta_0/2(\omega-\omega_0)^2$ with a large enough chirp rate $\beta_0$  can lead to robust control \cite{epjd,ejl,arXiv:1701.03541v2,pra:32:3435}.  By incorporating  a smooth  filtering function of frequency  into the optimization algorithm,
 our numerical simulations  show that  optimal solutions towards remarkable robustness  also turn out to be a quadratic spectral phase function, whereas optimal solutions found by the algorithm  in the absence of this  filtering function  are far from being  robust. From quantum optimal control point of view, our results  provide a new approach to search optimal robust solutions, yielding the high-fidelity value of a cost functional while resisting  the control field fluctuations. We further examine the SPOO algorithm in a three-level atomic rubidium, for which optimal robust solutions  are unknown before performing the optimization. Two different  schemes  of quantum state transfer are discussed for achieving optimal robust population inversion between levels with either allowed or forbidden electric dipole transitions.  
 \\ \indent
 The rest of the paper is organized as follows. In Sec. \ref{method}, we describe the details of the SPOO algorithm in the frame of FDQOCT. Numerical simulations and discussion are demonstrated in Sec. \ref{red} for the two-level quantum system  and three-level atomic rubidium. Our results
are briefly summarized in Sec. \ref{conc}.
\section{Theoretical Methods}\label{method}
Consider a general  system consisting of $N$ quantum states $|n\rangle$ with eigenenergies $|E_n\rangle$ ($n=1\cdots$, $N$). The total Hamiltonian operator $\hat{H}(t)$ of the quantum system in interaction with the temporal field $\mathcal{E}(t)$ can be described  by  $\hat{H}(t)=\hat{H}_0-\hat{\mu}\mathcal{E}(t)$, where $\hat{H}_0=\Sigma_{n=1}^NE_n|n\rangle\langle n|$ is the field-free Hamiltonian operator  and $\hat{\mu}$ denotes the dipole operator. The time-dependent evolution of the quantum system initially in state $|i\rangle$ at time $t_i$ is described by  the wave function $|\Psi(t)\rangle=\hat{U}(t,t_i)|i\rangle$.  The corresponding unitary evolution operator $\hat{U}(t,t_i)$ is governed by the time-dependent Schr\"{o}dinger equation,
\begin{eqnarray}
i\hbar\frac{\partial \hat{U}(t,t_i)}{\partial t}=\hat{H}(t)\hat{U}(t,t_i), \ \ \ \hat{U}(t_i,t_i)\equiv \mathbb{I}.\label{sheq}
\end{eqnarray}
To find an optimal control field $\mathcal{E}(t)$ in the ultrafast time scales, the FDOQCT has been formulated  previously in Ref. \cite{pra:93:033417} for optimizing the complex function $\boldsymbol{E}(\omega)$ subject to multiple equality constraints, which was further developed in Ref. \cite{pra:inpress} leading to a multi-objective frequency domain optimization algorithm.
In the following, we will describe the details of how employ the FDQOCT to optimize the real spectral phase function $\phi(\omega)$ of an ultrafast control field subject to multiple equality constraints. As a result, a multiple-equality-constraint SPOO algorithm is used  for maximizing a cost functional $P_{i\rightarrow f}(t_f)=|\langle f|\Psi(t_f)\rangle|^2$, i.e., the probability of quantum state transfer from the initial state $|i\rangle$ to the final state $|f\rangle$  at the final time $t_f$.\\ \indent
As demonstrated in the FDOQCT \cite{pra:93:033417},  a dummy variable $s\geq0$ is employed to record the changes of the spectral phase $\phi(\omega)$  and the cost functional $P_{i\rightarrow f}(t_f)$ with $\phi(s, \omega)$ and $P_{i\rightarrow f}(s, t_f)$. Note that the parameter $s$ introduced in the FDQOCT is not a  physical variable of the laser field,
which is used  for formulating the optimization algorithm by first-order differential equations  \cite{jcp:123:134104,pra:72:023416}.  As $s$ increases, updating the spectral phase from $\phi(s, \omega)$ to $\phi(s+\delta s, \omega)$ that increases  the value of $P_{i\rightarrow f}(t_f)$ (i.e., $P_{i\rightarrow f}(s+\delta s,t_f)-P_{i\rightarrow f}(s,t_f)\geq0$) can be written using the chain rule as
 \begin{eqnarray}\label{obj}
g_0(s)\equiv\frac{d P_{i\rightarrow f}(s,t_f)}{d s}=\int_{0}^{\infty}\frac{\delta
P_{i\rightarrow f}(s,t_f)}{\delta\phi(s,\omega)}\frac{\partial\phi(s,\omega)}{\partial
s}d\omega\geq0.
\end{eqnarray}
If there are no any constraints on the optimal control fields, Eq. (\ref{obj}) can be satisfied  by integrating the following  equation
\begin{eqnarray} \label{free}
\frac{\partial\phi(s,\omega)}{\partial s}=\frac{\delta
P_{i\rightarrow f}(s,t_f)}{\delta\phi(s,\omega)}. \label{uncons}
\end{eqnarray}
In this work, we further impose two equality external constraints
\begin{eqnarray}\label{constraint1}
g_1(s)\equiv\frac{d\mathcal{E}(s,t_i)}{d s}=\int_{0}^{\infty}\frac{\delta
\mathcal{E}(s,t_i)}{\delta\phi(s,\omega)}\frac{\partial\phi(s,\omega)}{\partial
s}d\omega=0,
\end{eqnarray}
and
\begin{eqnarray}\label{constraint2}
g_2(s)\equiv\frac{d\mathcal{E}(s,t_f)}{d s}=\int_{0}^{\infty}\frac{\delta
\mathcal{E}(s,t_f)}{\delta\phi(s,\omega)}\frac{\partial\phi(s,\omega)}{\partial
s}d\omega=0,
\end{eqnarray}
on the control fields during the whole optimization, which are often considered in quantum optimal control simulations  for practical implementation \cite{jpb:4:R175,prl:111:050404,pra:96:022309,njp18,prl:101:073002}. Optimal solutions of
$\partial\phi(s,\omega)/\partial s$ that simultaneously satisfy Eqs.
(\ref{obj}), (\ref{constraint1}) and (\ref{constraint2}) can be expressed as
\begin{eqnarray}\label{control field1}
\frac{\partial \phi(s,\omega)}{\partial
s}=g_0(s)\int_{0}^{\infty}S(\omega'-\omega)\sum_{\ell=0}^2
\left[\Gamma^{-1}\right]_{0\ell}
\frac{\delta \mathcal{Q}_{\ell}}{\delta\phi(s, \omega)}d\omega',\nonumber \\
\end{eqnarray}
where $\mathcal{Q}_0=P_{i\rightarrow f}(s, t_f)$, $\mathcal{Q}_1=\mathcal{E}(s,t_i)$ and $\mathcal{Q}_2=\mathcal{E}(s, t_f)$. A frequency domain convolution filtering function $S(\omega'-\omega)$ in Eq. (\ref{control field1}) is introduced  to  locally average the inputs $\delta \mathcal{Q}_{\ell}/\delta\phi(s, \omega)$.  In our simulations,  we take a normalized Gaussian function of frequency as the convolution filtering
 \begin{eqnarray}\label{filter}
S(\omega'-\omega)=\exp\Bigg[-\frac{4\ln2(\omega'-\omega)^2}{\sigma^2}\Bigg],
\end{eqnarray}
where $\sigma$ is the bandwidth. Although such a  convolution filtering has been involved in the FDQOCT \cite{pra:93:033417,pra:inpress}, we did not highlight its virtue for obtaining
robust solutions.
$\Gamma$ is a $3\times3$ symmetric matrix composed of the elements
\begin{eqnarray}\label{gama}
\Gamma_{\ell\ell'}=\int_{0}^{\infty}\delta
\mathcal{Q}_\ell/\delta \phi(s,\omega)\int_{0}^{\infty}S(\omega'-\omega)\delta
\mathcal{Q}_{\ell'}/\delta \phi(s,\omega')d\omega'd\omega.
\end{eqnarray}
By inserting Eq. (\ref{control field1}) into  Eqs.
(\ref{obj}), (\ref{constraint1}) and (\ref{constraint2}), we can verify that
\begin{eqnarray}
g_{\ell'}(s)&=&g_0(s)\int_{0}^{\infty}\delta\mathcal{Q}_{\ell'}/\delta \phi(s,\omega)\int_{0}^{\infty}S(\omega'-\omega)\sum_{\ell=0}^2
\left[\Gamma^{-1}\right]_{0\ell}\frac{\delta\mathcal{Q}_\ell}{\delta \phi(s,\omega)}d\omega'd\omega\nonumber \\
&=&g_0(s)\sum_{\ell=0}^M
\left[\Gamma^{-1}\right]_{0\ell}\Gamma_{\ell\ell'}\nonumber \\
&=&g_0(s)\delta_{0\ell'} \ \ \ \ell'=0,1,2. \label{sm1}
\end{eqnarray}
That is, all requirements from   Eqs.
(\ref{obj}), (\ref{constraint1}) and (\ref{constraint2}) are satisfied during the optimization simultaneously if the spectral phase is updated  with the algorithm Eq. ({\ref{control field1}}).    \\ \indent
The SPOO algorithm in Eq. (\ref{control field1}) is independent of the dimension
of Hamiltonian, ensuring its applicability to complex multi-level quantum systems. To perform this SPOO algorithm, the gradients $\delta\mathcal{Q}_1/\delta\phi(s,\omega)$ and $\delta\mathcal{Q}_2/\delta\phi(s,\omega)$  can be analytically given by
\begin{eqnarray} \label{de}
\frac{\partial \mathcal{E}(s,t)}{\partial \phi(s,\omega)}=\mathcal{A}(\omega)\sin[\omega t-\phi(s, \omega)]
\end{eqnarray}
with $t=t_i$ and $t_f$. The gradient $\delta\mathcal{Q}_0/\delta\phi(s,\omega)$ is computed by
\begin{eqnarray}
\frac{\delta \mathcal{Q}_0}{\delta \phi(s,\omega)}=\int_{-\infty}^{\infty}\frac{\delta
\mathcal{Q}_0}{\delta\mathcal{E}(s,t)}\frac{\partial \mathcal{E}(s,t)}{\partial \phi(s,\omega)}dt,
\end{eqnarray}
in which $\partial \mathcal{E}(s,t)/\partial \phi(s,\omega)$ has been derived in Eq. (\ref{de}), and the gradient  $\delta\mathcal{Q}_0/\delta\mathcal{E}(s,t)$ is calculated by \cite{pra:93:053418}
\begin{eqnarray}
 \frac{\delta \mathcal{Q}_0}{\delta \mathcal{E}(s,t)}&=&-2\mathrm{Im}\left\{\langle i|\hat{U}^\dagger(t_f, t_i)|f\rangle\langle f|\hat{U}(t_f, t_i)\hat{U}^\dagger(t, t_i)\hat{\mu}\hat{U}(t, t_i)|i\rangle\right\}.
\end{eqnarray}
\\ \indent
In our simulations,  the spectral amplitude $\mathcal{A}(\omega)$ is fixed with a Gaussian frequency distribution centered at the frequency $\omega_0$,
\begin{eqnarray}
\mathcal{A}(\omega)=\mathcal{E}_0\sqrt{\frac{1}{2\pi\Delta\omega^2}}\exp\Bigg[-\frac{(\omega-\omega_0)^2}{2\Delta\omega^2}\Bigg],
\end{eqnarray}
where $\Delta\omega$ is the frequency bandwidth, and $\mathcal{E}_0$ is the peak field strength. Equation (\ref{sheq}) is firstly solved by using  a temporal field $\mathcal{E}(s_0,t)$  with an initial guess of the spectral phase $\phi(s_0,\omega)$, and then the generated wavefunction $|\Psi(t)\rangle$ is used to calculate $\delta\mathcal{Q}_0/\delta\phi(s_0,\omega)$. The first-order differential equation  (\ref{control
field1}) is solved (e.g., by using the Euler method) to obtain the first updated spectral phase $\phi(s_1=s_0+\delta s, \omega)=\phi(s_0, \omega)+\delta s(\partial \phi(s_0,\omega)/\partial s)$, which combined with the fixed spectral amplitude $\mathcal{A}(\omega)$ is used to calculate the first updated control field  $\mathcal{E}(s_1,t)$. Equation (\ref{sheq}) is further solved by using the updated field  $\mathcal{E}(s_1,t)$, which will increase the cost functional  of $\mathcal{Q}_0$  (i.e., $P_{i\rightarrow f}(t_f)$) as compared with that by the initial  field $\mathcal{E}(s_0,t)$. By repeating the step $s_0$ to the step $s_1$, the spectral phase is iteratively updated from $\phi(s_1,\omega)$ to $\phi(s_2=x_1+\delta s,\omega), \cdots, \phi(s_n,\omega)$ until
$\mathcal{Q}_0$, i.e., $P_{i\rightarrow f}(t_f)$,  converges to the desired precision.
\\ \indent
\section{Results and Discussion}\label{red}
\begin{figure}[!t]\centering
\resizebox{0.6\textwidth}{!}{%
  \includegraphics{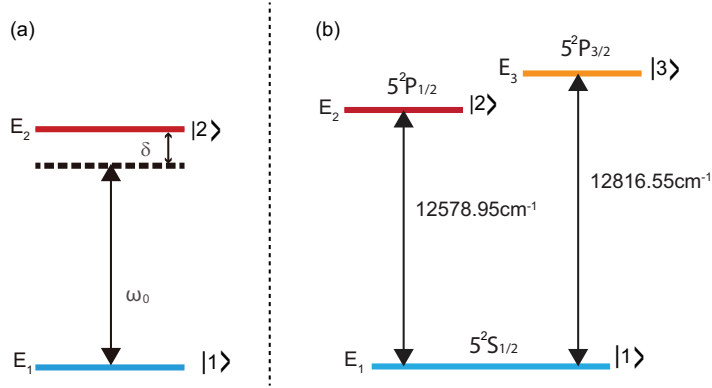}
} \caption{Schematic energy level structures of a two-level quantum system (a) and a three-level atomic rubidium (b). For two-level simulations, the energies of the two states $|1\rangle$ and $|2\rangle$ in (a) are fixed at $E_1=0$ and $E_2=12500$ cm$^{-1}$, and $\delta$ denotes a single-photon detuning between the center frequency $\omega_0$ and the transition frequency $\omega_{12}=(E_2-E_1)/\hbar$. For three-level simulations, the three states $|1\rangle$, $|2\rangle$ and $|3\rangle$ correspond to the energy levels $5^2S_{1/2}$, $5^2P_{1/2}$ and $5^2P_{3/2}$ in atomic rubidium with the energies $E_1=0$, $E_2=12578.95$ and $E_3=12816.55$ cm$^{-1}$, respectively.}
\label{fig1}
\end{figure}
In the following, we will perform the  SPOO algorithm in  an abstract two-level quantum system and in  a three-level atomic rubidium, respectively. For two-level simulations, the energies of the two states $|1\rangle$ and $|2\rangle$ in Fig. \ref{fig1} (a) are fixed at $E_1=0$ and $E_2=12500$ cm$^{-1}$, and $\delta$ denotes a single-photon detuning between the center frequency $\omega_0$ and the transition frequency $\omega_{12}=(E_2-E_1)/\hbar$.  The transition dipole moments between the states $|1\rangle$ and $|2\rangle$ are chosen to be $\mu_{12}=\mu_{21}=1.0$ a.u. without loss of generality.
For three-level simulations, the three states $|1\rangle$, $|2\rangle$ and $|3\rangle$ in Fig. \ref{fig1} (b) correspond to the energy levels $5^2S_{1/2}$, $5^2P_{1/2}$ and $5^2P_{3/2}$ for atomic rubidium-87 with $E_1=0$, $E_2=12578.95$ and $E_3=12816.55$ cm$^{-1}$, respectively. The transition dipole moments between states are taken as  $\mu_{12}=\mu_{21}=2.9931$ and $\mu_{13}=\mu_{31}=4.2275$ a.u., and $\mu_{23}=\mu_{32}$ are set to be zero for describing the electric dipole forbidden atomic transitions between the states $|2\rangle$ and $|3\rangle$.
\subsection{Application to an abstract two-level quantum system}
\begin{figure}[!t]\centering
\resizebox{0.6\textwidth}{!}{%
  \includegraphics{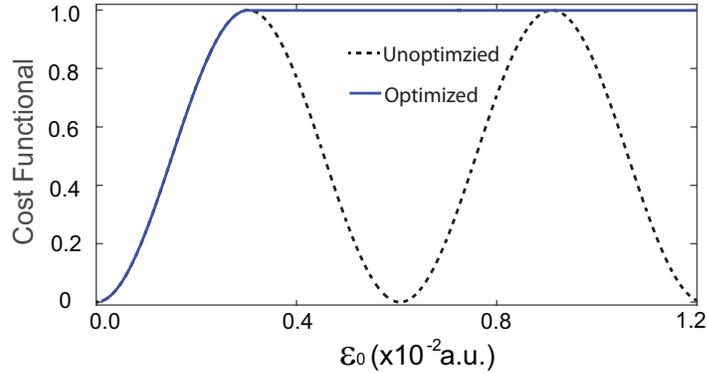}
} \caption{Two-level simulations: The cost functional $P_{1\rightarrow2}(t_f)$ as a function of the field strength $\mathcal{E}_0$ by using the transform limited (dashed lines) and spectral phase optimized (solid line) pulses.  The optimized calculations are accomplished by optimizing the spectral phase  without involving the filtering function $S(\omega'-\omega)$ in  Eq. (\ref{control field1}) while fixing the spectral amplitude $\mathcal{A}(\omega)$ as the corresponding transform limited pulse.}
\label{fig2}
\end{figure}
 We first perform the two-level simulations by using  a zero spectral phase and fixing the center frequency $\omega_0$ in resonance with the transition frequency $\omega_{12}$, for which the temporal control field in Eq. (\ref{phase-only}) corresponds to a transform limited pulse, $\mathcal{E}(t)=\mathcal{E}_0\exp(-t^2/2\tau_0^2)\cos\omega_0t$ with a duration of $\tau_0=1/\Delta\omega$.  In our simulations, the total propagation time is taken as $T=t_f-t_i=100\tau_0=1000$ fs, which is discretized with $3.2\times10^5$ uniform time steps. The quantum system is initially in the state $|1\rangle$, and finally is expected to be in the state $|2\rangle$.
  We scan the field strength $\mathcal{E}_0$ with different values to solve the time-dependent Schr\"{o}dinger equation (\ref{sheq}), and the corresponding $P_{1\rightarrow2}(t_f)$ as a function of $\mathcal{E}_0$ is plotted in Fig. \ref{fig2} (dashed lines).  Since the exact resonant condition (i.e., $\delta=0$) is satisfied, the transition probability $P_{1\rightarrow2}(t_f)$ can be analytically expressed as   \cite{scully} $P_{1\rightarrow2}(t_f)=\sin^2(A(t_f)/2)$ with $A(t_f)=\mathcal{E}_0\mu_{12}\int_{t_i}^{t_f}\exp(-t^2/2\tau_0^2)/2dt\propto\mathcal{E}_0$, which as shown in Fig. \ref{fig2} oscillates between 0 and 1 depending on the value of $\mathcal{E}_0$. Complete population transfer occurs for $A(t_f)=(2q+1)\pi$ (i.e., odd-$\pi$ pulse) and complete population return takes place for $A(t_f)=2q\pi$ (i.e., even-$\pi$ pulse), where $q$ is an integer.  This phenomenon is either called Rabi oscillation or Rabi flopping.
   The odd-$\pi$ pulse provides an approach for achieving complete population transfer, whereas as demonstrated in Fig. \ref{fig2} the obtained population of $P_{1\rightarrow2}(t_f)$ is not robust with respect to  the fluctuation of $\mathcal{E}_0$. To explore whether there are optimal robust solutions, the  SPOO algorithm is first employed  without using the  filtering function $S(\omega'-\omega)$. Optimal spectral phases with different values of $\mathcal{E}_0$ are obtained, and the corresponding $P_{1\rightarrow2}(t_f)$  as a function of $\mathcal{E}_0$ is plotted in Fig. \ref{fig2} (solid line). It is clear that $P_{1\rightarrow2}(t_f)$ with a high-fidelity is possible in a large range of $\mathcal{E}_0$ (i.e. $A(t_f)\geq\pi$).  That is, there exist  many  optimal spectral phases $\phi(\mathcal{E}_0, \omega)$, which are able to obtain the observable value  of  $P_{1\rightarrow2}(t_f)>0.9999$ with an admissible error  $<10^{-4}$.  \\ \indent
  \begin{figure}[!t]\centering
\resizebox{0.6\textwidth}{!}{%
  \includegraphics{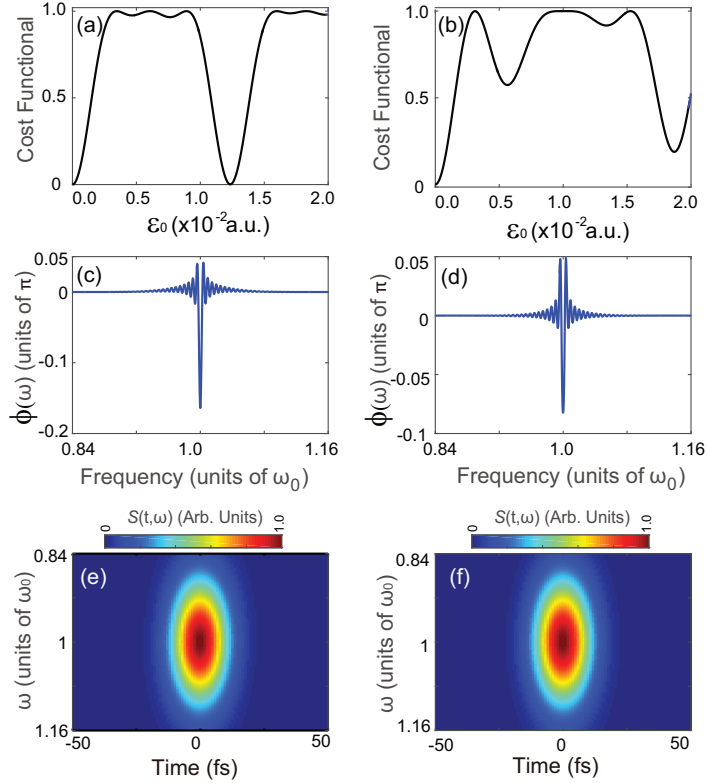}
} \caption{Two-level simulations: Robustness with respect to the field strength variations for two different optimal  spectral phases. (a) and (b) The cost functional $P_{1\rightarrow2}(t_f)$ as a function of $\mathcal{E}_0$, (c) and (d) the corresponding spectral phases, and  (e) and (f) the  time and frequency resolved  distributions of the control fields. The two optimized spectral phases are obtained from  the unfiltered optimizations at $\mathcal{E}_0=0.6\times10^{-2}$ (left panels) and $1.0\times10^{-2}$ a.u. (right panels).}
\label{fig3}
\end{figure}
To see whether the optimized spectral phases are robust,  two different optimized spectral phases obtained with  $\mathcal{E}_0=0.6\times10^{-2}$  and $1.0\times10^{-2}$ a.u. that lead to $P_{1\rightarrow2}(t_f)>0.9999$  in Fig. \ref{fig2} are examined, respectively. The corresponding robustness of  $P_{1\rightarrow2}(t_f)$ against the fluctuation of $\mathcal{E}_0$  is plotted in Figs. \ref{fig3} (a) and (b). As can be seen from Figs. \ref{fig3} (c) and (d), the optimized spectral phases are mainly modulated in a small region around the center frequency $\omega_0$, showing strong oscillations in the values.  Figures \ref{fig3} (e) and (f) plot  the corresponding time and frequency resolved distributions of the control fields, which  are almost unchanged as compared with the corresponding transform limited pulses. That is, the optimized spectral phases only lead to a small modulation on the initial temporal field, but such a slight modification of the initial guess (i.e.,  a transform limited pulse) is capable of  maximizing  $P_{1\rightarrow2}(\mathcal{E}(\cdot))$. We also examine other optimized spectral phases (not shown here), showing similar behaviours as Fig. \ref{fig3}. Thus, the SPOO algorithm in the absence of the filtering function can be used to find optimal solutions, but they are far from being robust. \\ \indent
\begin{figure}[!t]\centering
\resizebox{0.6\textwidth}{!}{%
  \includegraphics{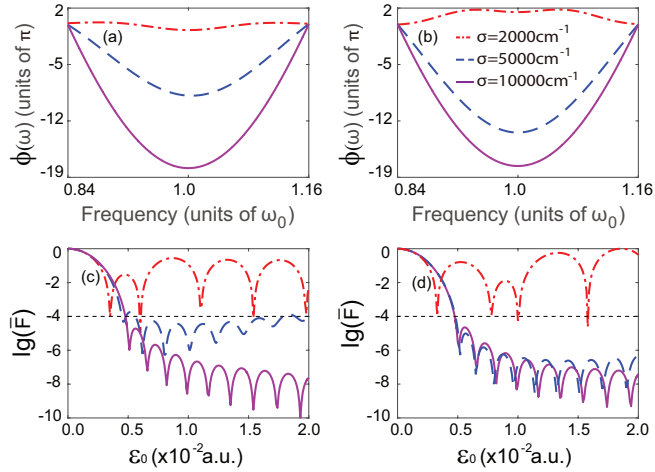}
} \caption{Two-level simulations: The effect of the parameter $\sigma$ on the optimal spectral phases and  infidelity $\bar{F}=1-P_{1\rightarrow2}(t_f)$.
 The optimal spectral phases for  three different values of $\sigma=2.0\times10^{3}, 5.0\times10^{3}$ and $1.0\times10^{4}$ cm$^{-1}$ are calculated with (a) $\mathcal{E}_0=0.6\times10^{-2}$  and (b) $1.0\times10^{-2}$ a.u., and the corresponding infidelity $\bar{F}=1-P_{1\rightarrow2}(t_f)$ in decimal logarithmic scale as a function of $\mathcal{E}_0$ is plotted in (c) and (d).   The dashed horizontal lines correspond to the infidelity $\bar{F}=1.0\times10^{-4}$.}
\label{fig4}
\end{figure}
We  apply the SPOO algorithm with the filtering function to the above problem. To show the role of the parameter $\sigma$ in the filtering function,  we examine the optimization algorithm at the two different field strengths $\mathcal{E}_0=0.6\times10^{-2}$  and $1.0\times10^{-2}$ a.u. as used in the above simulations  with different values of $\sigma=2.0\times10^{3}, 5.0\times10^{3}$ and $1.0\times10^{4}$ cm$^{-1}$, and the corresponding optimal spectral phases are plotted in Figs. \ref{fig4} (a) and (b). We can find that the optimal spectral phases are very sensitive to the parameter $\sigma$, indicating that there are still many optimal solutions with the same spectral amplitude $\mathcal{A}(\omega)$.
 To clearly view the dependance of the robustness on the parameter $\sigma$ with these optimal spectral phases,  Figs. \ref{fig4} (c) and (d) show the corresponding infidelity $\bar{F}=1-P_{1\rightarrow2}(t_f)$ in decimal logarithmic scale as a function of $\mathcal{E}_0$, which usually is used to assess a quantum state transformation error in  quantum information science. We  find that the infidelity is dramatically  decreased to  the  admissible error below $10^{-4}$ with the value of $\sigma\geq5.0\times10^{3}$ cm$^{-1}$, where the corresponding  optimal spectral phases in Figs. \ref{fig4} (a) and (b) show a smooth change in $\omega$ without strong oscillations. \\ \indent
 \begin{figure}[!t]\centering
\resizebox{0.6\textwidth}{!}{%
  \includegraphics{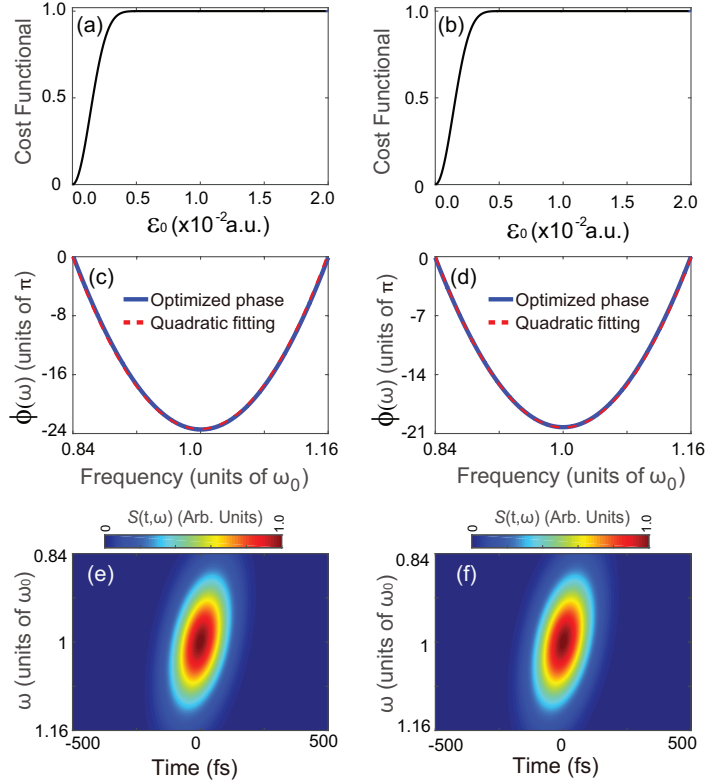}
} \caption{Two-level simulations: Robustness with respect to the field strength variations for two different optimal  spectral phases. (a) and (b) The cost functional $P_{1\rightarrow2}(t_f)$ as a function of $\mathcal{E}_0$, (c) and (d) the corresponding spectral phases, and  (e) and (f) the  time and frequency resolved  distributions of the control fields. The two optimized spectral phases (blue solid lines) are obtained from  the filtered optimizations at $\mathcal{E}_0=0.6\times10^{-2}$ (left panels) and $1.0\times10^{-2}$ a.u. (right panels), which with a constant shift can be fitted (red dashed lines) by using a quadratic function of $\beta_0/2(\omega-\omega_0)^2$  with a chirp rate $\beta_0=898$ fs$^2$ in (c) and 1044 fs$^2$ in (d).}
\label{fig5}
\end{figure}
 Figure \ref{fig5} displays the same simulations as Fig. \ref{fig3} by involving the filtering function with the parameter of $\sigma=2.0\times10^4$ cm$^{-1}$.   The robustness against  $\mathcal{E}_0$ is examined in Figs. \ref{fig5} (a) and (b) with the two optimized spectral phases in Figs. \ref{fig5} (c) and (d), showing remarkable  robustness to large changes in the field strength $\mathcal{E}_0$.  It is interesting to note that  optimized spectral phases with a constant shift can be fitted very well using  a quadratic spectral phase $\beta_0/2(\omega-\omega_0)^2$ with a chirp rate $\beta_0$, which  clearly introduces a time-dependent frequency distribution of the field, as shown in Figs. \ref{fig5} (e) and (f) and prolongs  the durations $\tau$ of the optimized pulses much longer than $\tau_0=10$ fs. As a result, shaping the spectral phase will reduce  the peak intensity of the optimized temporal electric fields as compared with the transform-limited pulse (see also details in Appendix).  The underlying mechanism of
generating robust control of  a two-level quantum system with such a  linear frequency-chirped pulse can be understood well in terms of the noncrossing adiabatic representation (see details in Appendix). As demonstrated in Figs. \ref{fig6} (a) and (b),  the evolution of quantum systems is almost completely  in the adiabatic ground state $|-\rangle$ during the whole quantum control process. As a result,  the quantum system in the diabatic representation  smoothly evolves from the initial state $|1\rangle$ to the final state $|2\rangle$, as shown in Figs. \ref{fig6} (c) and (d). That is,  an adiabatic pulse is designed by optimizing the spectral phase of the initial nonadiabatic ultrafast pulse, which leads to a remarkable robustness against  the variations of the field strength.
 \\ \indent   
\begin{figure}[!t]\centering
\resizebox{0.6\textwidth}{!}{%
  \includegraphics{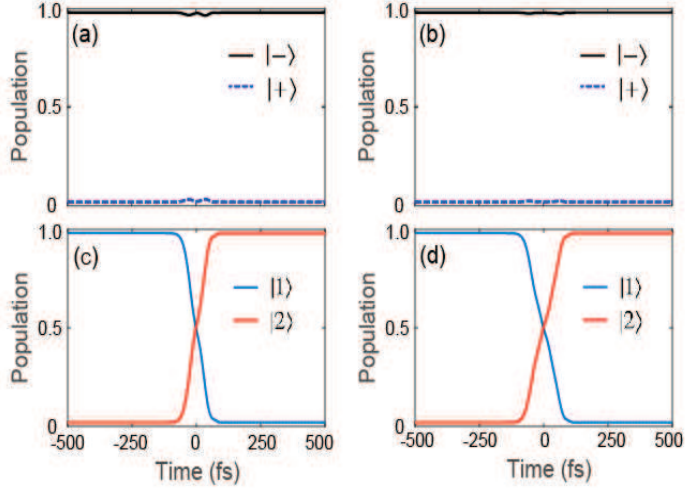}
} \caption{Two-level simulations: Time evolution of the qubit control  in (a) and (b) adiabatic and (c) and (d) diabatic states  for the two filtered spectral phase-optimized pulses. The spectral amplitude $A(\omega)$ is fixed at (left panels) $\mathcal{E}_0=0.6\times10^{-2}$ and (right panels) $1.0\times10^{-2}$ a.u.}
\label{fig6}
\end{figure}
Optimal control theory combined with adiabatic theorem has been applied previously for the time-domain design of adiabatic microsecond radio-frequency pulses
by explicitly incorporating an adiabaticity term into the cost functional \cite{mrm:36:401}. The present method is performed in the frequency domain for the design of  \emph{ultrafast} pulses without using such an adiabaticity term as the part of control objective, whereas as demonstrated in the two-level simulations the spectral filtering function with a large enough value of $\sigma$  ``implicitly'' enforces optimal solutions towards adiabatic pulses, providing a new approach to  search  for optimal robust solutions.
\subsection{Application to a three-level atomic rubidium}
\begin{figure}[!t]\centering
\resizebox{0.6\textwidth}{!}{%
  \includegraphics{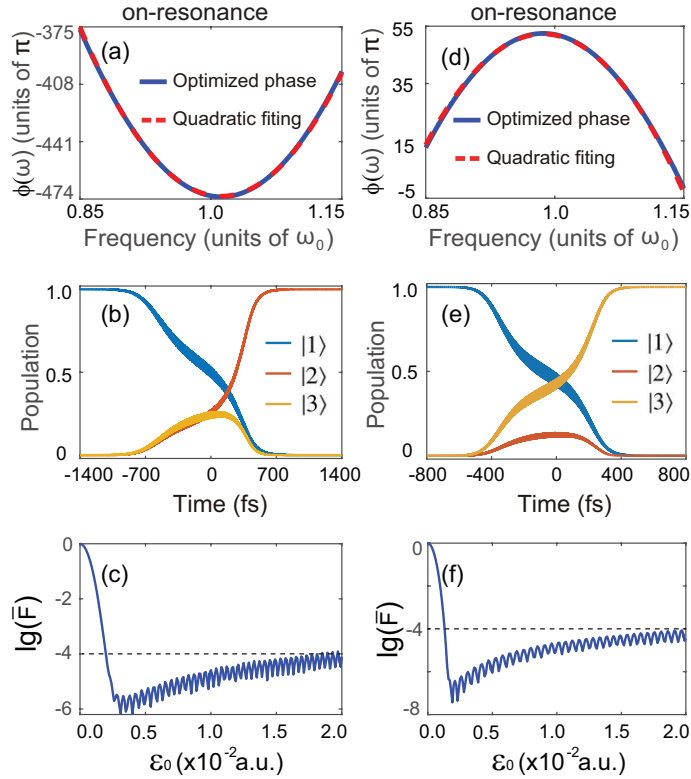}
} \caption{Three-level simulations for on-resonance excitations: The center frequency $\omega_0$ is fixed in resonance with the transition frequency  $\omega_{12}=(E_2-E_1)/\hbar$ in (left panels) and  $\omega_{13}=(E_3-E_1)/\hbar$ (right panels). (a) and (d): The optimal spectral phases (blue solid lines) are obtained with $\mathcal{E}_0=0.8\times10^{-2}$ a.u. and $\sigma=1.8\times10^4$ cm$^{-1}$, which with a constant shift are fitted (red dashed lines) by using a quadratic function $\beta_0/2(\omega-\omega_c)^2$ (a) with $\omega_c=12727.39$ cm$^{-1}$ and $\beta_0=4161$ fs$^2$ and (d) with $\omega_c=12644.675$ cm$^{-1}$ and $\beta_0=-2235$ fs$^2$. (b) and (e): The  time-dependent population transfer among the three states. (c) and (f): The  infidelity $\bar{F}=1-P_{1\rightarrow2}(t_f)$  and $\bar{F}=1-P_{1\rightarrow3}(t_f)$  in decimal logarithmic scale as a function of $\mathcal{E}_0$.  The dashed horizontal lines correspond to the infidelity $\bar{F}=1.0\times10^{-4}$. }
\label{fig7}
\end{figure}
\subsubsection{On-resonance case}
We apply  the SPOO algorithm with the filtering function to  the three-level atomic rubidium, as shown in Fig. \ref{fig1} (b). We first examine two different on-resonance excitation schemes from state $|1\rangle$ by fixing the center frequency $\omega_0$ in resonance with the transition frequency  $\omega_{12}=(E_2-E_1)/\hbar$ or $\omega_{13}=(E_3-E_1)/\hbar$, and the corresponding final state is $|2\rangle$ or $|3\rangle$, respectively. Figure \ref{fig7} shows  the corresponding simulations with $\mathcal{E}_0=0.8\times10^{-2}$ a.u. and $\sigma=1.8\times10^4$ cm$^{-1}$. The optimal spectral phase in Fig. \ref{fig7} (a) or in Fig. \ref{fig7} (d) with a constant shift  in the value can  be fitted very well by using a quadratic function  $\beta_0/2(\omega-\omega_c)^2$  centered at the frequency $\omega_c$.   Figure \ref{fig7} (c) or (e) shows the corresponding infidelity $\bar{F}=1-P_{1\rightarrow2}(t_f)$ or $\bar{F}=1-P_{1\rightarrow3}(t_f)$ in decimal logarithmic scale as a function of $\mathcal{E}_0$. A high fidelity  with the  admissible error of  $\bar{F}<10^{-4}$ is observed in a large range of $\mathcal{E}_0$. We also examine this on-resonance case with other values of $\mathcal{E}_0$, and the similar behaviours are observed with a large value of $\sigma$.  We can see that the frequency distribution of the optimal spectral phase is not symmetric to  the center frequency $\omega_0$, and therefore a frequency detuned  quadratic spectral phase function is found as an optimal robust solution.
As can be seen from Fig. \ref{fig7} (b) or from Fig. \ref{fig7} (e), the state $|3\rangle$ or $|2\rangle$  is clearly involved in the population transfer process, whereas the final population in this state is efficiently suppressed to a very low value of $<10^{-4}$.  Furthermore, we can find in Fig. \ref{fig7} (c) or in Fig. \ref{fig7} (f) that increasing the laser pulse strength causes a rising trend of infidelity. This can also be attributed to the existence of the state $|3\rangle$ or $|2\rangle$, as the frequency bandwidth of the  pulse is broad enough to excite both the excited states $|2\rangle$ and $|3\rangle$ with an energy
separation of $E_3-E_2\approx238$ cm$^{-1}$.  Based on these considerations, the optimal robust pulse does not construct an adiabatic passage between the initial state  and the final state, whereas it is able to achieve a high-fidelity of $P_{1\rightarrow2}(t_f)>0.9999$ or $P_{1\rightarrow3}(t_f)>0.9999$ against the fluctuation of  $\mathcal{E}_0$.\\ \indent
For such a three-level atomic rubidium, we have noticed that the three levels involved can also be approximated to form a two-level system with the states $|1\rangle$ and $|2\rangle$ by choosing a set of optimal parameters  ($\Delta\omega, \delta$, $\beta_0$) \cite{arXiv:1701.03541v2}, so that the effect of the state $|3\rangle$ on quantum state transfer can be sufficiently suppressed. That is, we may also form a two-level system by setting the spectral amplitude $\mathcal{A}(\omega)$ with the frequency  bandwidth $\Delta\omega$ small enough to overlap significantly only the final state of interest \cite{epjd}. As a result, optimal robust solutions  will turn out to be the  two-level case with adiabatic pulses. That is, the dynamics of a multi-level quantum system is reduced to an effective two-level model.  \\ \indent
\begin{figure}[!t]\centering
\resizebox{0.6\textwidth}{!}{%
  \includegraphics{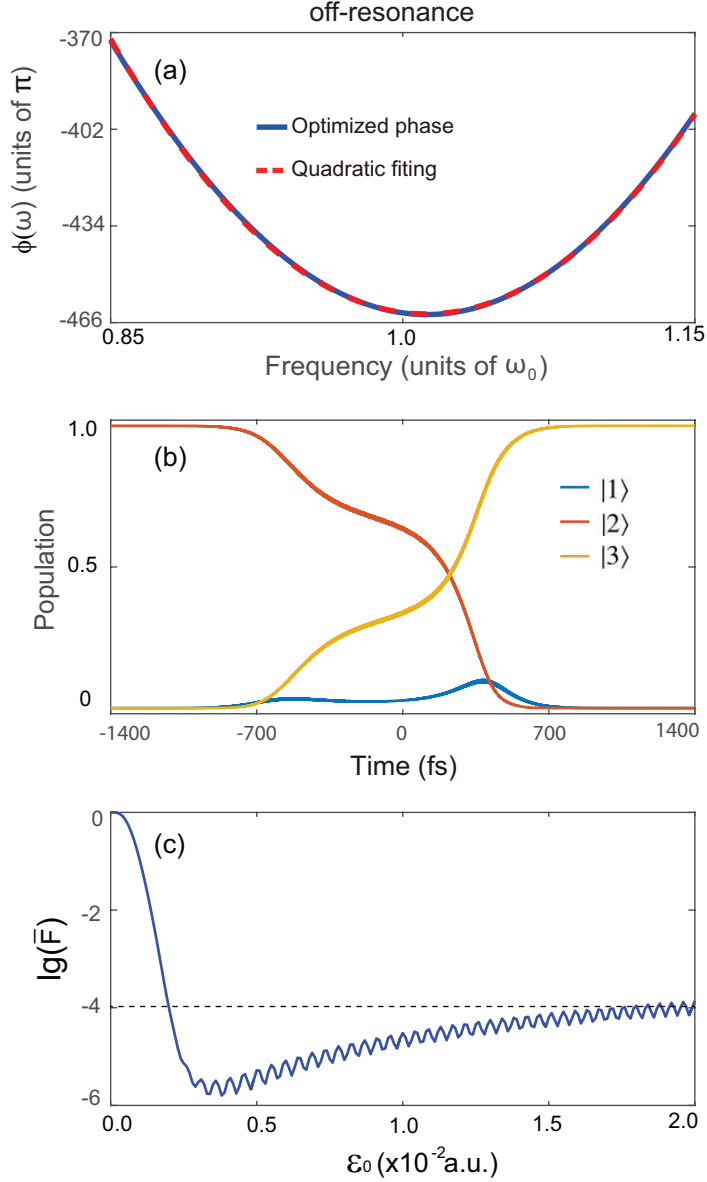}
} \caption{Three-level simulations for off-resonance excitations:  The system is initially in the sate $|2\rangle$ with the state $|3\rangle$ as the target, and the center frequency $\omega_0$ is fixed in resonance with the transition frequency  $\omega_{12}=(E_2-E_1)/\hbar$.  (a): The optimal spectral phase (blue solid line) is obtained with $\mathcal{E}_0=0.8\times10^{-2}$ a.u. and $\sigma=1.8\times10^4$ cm$^{-1}$, which with a constant shift is fitted (red dashed lines) by using a quadratic function $\beta_0/2(\omega-\omega_c)^2$ with $\omega_c=12735.2$ cm$^{-1}$ and $\beta_0=3884$ fs$^2$. (b): The time-dependent population transfer among the three states. (c): The corresponding infidelity $\bar{F}=1-P_{2\rightarrow3}(t_f)$ in decimal logarithmic scale as a function of $\mathcal{E}_0$. The dashed horizontal lines correspond to the infidelity $\bar{F}=1.0\times10^{-4}$. }
\label{fig8}
\end{figure}
\subsubsection{Off-resonance case}
Figure \ref{fig8} shows the off-resonance case in the three-level atomic rubidium with  $\mathcal{E}_0=0.8\times10^{-2}$ a.u. and $\sigma=1.8\times10^4$ cm$^{-1}$. For this off-resonance excitation, we assume that the initial state is $|2\rangle$, and  fix the center frequency $\omega_0$  in resonance with the transition frequency  $\omega_{12}=(E_2-E_1)/\hbar$.  An optimal spectral phase, as shown in Fig. \ref{fig8} (a), is found, which leads to the evolution of  the system from the initial state $|2\rangle$ to the final state $|3\rangle$.  The optimal spectral phase with a constant shift in the value can also be fitted very well by using  a quadratic function  $\beta_0/2(\omega-\omega_c)^2$. Although the direct transition between the states $|2\rangle$ and $|3\rangle$ is forbidden, the intermediate state $|1\rangle$ due to the broad  bandwidth of the pulse can be involved to construct  two (multi)-photon transition pathways. As can be seen from Fig. \ref{fig8} (b), the population is transferred to this intermediate state. Note that the population transfer processes involved  in this three-level system  are different from that by using the stimulated adiabatic Raman passage (STIRAP) scheme within  a V-type configuration system,  where the intermediate state is never populated  during the whole population transfer process \cite{rmp:89:015006,pra:79:023418}. Thus, this is not a completely adiabatic quantum state transfer from the initial state to the final state  by using the optimized pulse, whereas we can see that the population in the intermediate state $|1\rangle$ is not noticeable during the population transfer process. As a result, it is still able to lead to a very low error in a large range of $\mathcal{E}_0$. \\ \indent
Finally, we have an analysis for the center frequency detuning $\omega_c-\omega_0$ involved in the both on- and off-resonance simulations. We can  expand the optimal spectral phase around the center frequency $\omega_0$ by using the second-order Taylor expansion  $\phi(\omega)=\Sigma_{n=0}^{2}(\phi_n/n!)(\omega-\omega_0)^n$, where the first-order term is involved to lead to a shift of the spectral phase shape  in the frequency domain. According to the Fourier transform shift theorem, a linear
term in the spectral phase leaves the control  field envelope unchanged, and only  shifts the pulse in the time domain. As a result, the optimal solutions can also be understood as  the chirped pulses, which  transform an initial quantum state into a desired final state through an intermediate state by sweeping instantaneous frequency of laser pulse. 
\section{Conclusion}\label{conc}
In summary, we have demonstrated a theoretical study  to show how the spectral phase of an \emph{ultrafast} laser pulse can be shaped   to achieve optimal robust  control of quantum systems without shaping the  spectral amplitude of the laser pulse. A SPOO algorithm was established in the framework of FDQOCT, and was successfully applied for quantum state transfer in the abstract two-level quantum system  as well as in the three-level atomic rubidium.  By incorporating the filtering function into the optimization algorithm, the optimal spectral phases that lead to robust and high-fidelity quantum state transfer are found, and therefore we have shown an efficient approach  to  enforce optimal control algorithm in the frequency domain to extract optimal robust solutions. 
\\ \indent
 As this frequency domain optimization approach is in line with the current ultrafast pulse shaping technique commonly used in QOCEs,  this work together with optimization  algorithms may open a new  access to achieve optimal robust feedback control of quantum systems, for which both the spectral phase and amplitude of the ultrafast laser pulses can be used as the control variables.   This method in principle can be potentially applied to  more complex quantum control problems by either increasing the dimensionality of quantum systems or considering interactions of quantum systems with its environments, e.g.,  electron spins in diamond \cite{nature:532:77}, single molecules in  polymer hosts polymers \cite{nat:7:172}, and therefore we expect that it can be used in a robust way to optimize  specific pathways in chemical reactions and energy transfer channels in light-harvesting complexes, to achieve various quantum gates for quantum computing. 
     \\ \indent
 \section{Acknowledgment}
 Y.G. is partially  supported by the scholarship of Hunan Provincial Department of Education of China under grant No. 2015210, and by Hunan Provincial Natural
Science Foundation of China under grant No. 2017JJ2272.    C.C.S acknowledges the financial support by the Vice-Chancellor's Postdoctoral Research Fellowship of University of New South Wales (UNSW), Australia. D.D. acknowledges partial  supports  by the Australian Research Council under Grant No. DP130101658.
Y.G. also acknowledges the support and hospitality provided by UNSW Canberra
during his visit.

\section*{Appendix}
\appendix
\setcounter{section}{1}
The time-dependent electric field of a chirped pulse with a quadratic spectral phase takes the form \cite{jcp:134:164308}
 \begin{eqnarray}\label{chirp}
\mathcal{E}\left(t\right)=\mathcal{E}_{0}\mathrm{Re}\left\{ \sqrt{\frac{\tau_{0}^{2}}{\tau_{0}^{2}-i\beta_{0}}}\exp\left[-\frac{t^{2}}{2\tau^{2}}-i\left(\frac{\beta}{2}t+\omega_{0}\right)t\right]\right\},
\end{eqnarray}
where $\beta=\beta_{0}/(\tau_{0}^{4}+\beta_{0}^{2})$ and  $\tau=\tau_{0}\sqrt{1+\beta_{0}^{2}/\tau_{0}^{4}}$. By substituting the complex-valued $\sqrt{\tau_{0}^{2}/(\tau_{0}^{2}-i\beta_{0})}=fe^{-i\varphi}$ into Eq. (\ref{chirp}), the time-dependent electric field $\mathcal{E}\left(t\right)$  can be written
as
\begin{eqnarray}
\mathcal{E}\left(t\right)=\mathcal{E}'_{0}\left\{ \exp\left(-\frac{t^{2}}{2\tau^{2}}\right)\cos\left[\left(\frac{\beta}{2}t+\omega_{0}\right)t+\varphi\right]\right\}
\end{eqnarray}
 with an updated field strength $\mathcal{E}'_{0}=\mathcal{E}_{0}f$.  
 
Within  the rotating wave  approximation, the diabatic interaction Hamiltonian can be described by
\begin{eqnarray}
\hat{H}_{dia}=\frac{\hbar}{2}\left(
          \begin{array}{cc}
           -\Delta(t) & \Omega(t)\exp(i\varphi)\\
           \Omega(t)\exp(-i\varphi) & \Delta(t)\\
          \end{array}
        \right),
\end{eqnarray}
where $\Delta(t)=\delta-\beta t$ is the instantaneous detuning  and $\Omega(t)=-\mu_{12}\mathcal{E}'_{0}\exp(-t^2/2\tau^2)/\hbar$ is the Rabi frequency.  This leads to a modified Landau-Zener  model with a constant variation rate $\beta$ of the
energy difference \cite{landau,zener}, whereas the time-dependent diabatic coupling $\Omega(t)$ is involved. Clearly, the diabatic  energy levels in the absence of $\Omega(t)$ will take place an exact crossing when the energy level is swept.  In the presence of $\Omega(t)$,  however,  the adiabatic energy levels $E_{\pm}(t)=\pm\hbar\sqrt{\Omega^2(t)+\Delta^2(t)}/2$ obtained by diagonalizing $\hat{H}_{dia}$ will form an avoided crossing by slowly chirping the instantaneous frequency of the control field with a large enough chirp rate $\beta_0$ combined  with a large enough Rabi frequency $\Omega(t)$, i.e., the adiabatic condition of $|\dot{\vartheta}(t)|\ll\sqrt{\Delta^2(t)+\Omega^2(t)}$ is maintained   \cite{prl:100:103004,ejl,pra:93:023423}. The corresponding  adiabatic eigenstates can be given by  $|+\rangle=\sin\vartheta(t)|1\rangle e^{i\varphi}+\cos\vartheta(t)|2\rangle$ and $|-\rangle=\cos\vartheta(t)|1\rangle e^{i\varphi}-\sin\vartheta(t)|2\rangle$ with a mixing angle  $\vartheta(t)=\tan^{-1}(\Omega(t)/\Delta(t))/2$.     \\ \indent

\end{document}